\documentclass[prl,twocolumn,superscriptaddress,showpacs,nofootinbib]{revtex4}
\usepackage{graphicx}% Include figure files

\usepackage{bm}% bold math
\usepackage{amssymb}
\usepackage{amsmath}
\usepackage{amstext}
\usepackage{latexsym}

\newcommand{\nut}{\nu_t}

\begin{document}

\title{Structural defects in ion crystals by quenching the external potential: the inhomogeneous Kibble-Zurek mechanism}

\author{A. del Campo}
\affiliation{Institut f{\"u}r Theoretische Physik, Albert-Einstein Allee 11,
Universit{\"a}t Ulm, D-89069 Ulm, Germany}
\affiliation{QOLS, The Blackett Laboratory, Imperial College London, Prince Consort Road, SW7 2BW London, UK}
\author{G. De Chiara}
\affiliation{Grup d'{\`O}ptica, Departament de F{\'i}sica, Universitat Aut{\`o}noma
de Barcelona, E-08193 Bellaterra, Spain}
\affiliation{Grup de F\'isica Te\`orica, Departament de
F\'{i}sica, Universitat Aut\`{o}noma de Barcelona, E-08193
Bellaterra, Spain}
\author{Giovanna Morigi}
\affiliation{Grup d'{\`O}ptica, Departament de F{\'i}sica, Universitat Aut{\`o}noma
de Barcelona, E-08193 Bellaterra, Spain}
\affiliation{Theoretische Physik, Universit\"at des Saarlandes, D-66041 Saarbr\"ucken, Germany}
\author{M. B. Plenio}
\affiliation{Institut f{\"u}r Theoretische Physik, Albert-Einstein Allee 11,
Universit{\"a}t Ulm, D-89069 Ulm, Germany}
\affiliation{QOLS, The Blackett Laboratory, Imperial College London, Prince Consort Road, SW7 2BW London, UK}
\author{A. Retzker}
\affiliation{Institut f{\"u}r Theoretische Physik, Albert-Einstein Allee 11,
Universit{\"a}t Ulm, D-89069 Ulm, Germany}
\affiliation{QOLS, The Blackett Laboratory, Imperial College London, Prince Consort Road, SW7 2BW London, UK}

\def\d{{\rm d}}
\def\la{\langle}
\def\ra{\rangle}
\def\om{\omega}
\def\Om{\Omega}
\def\vep{\varepsilon}
\def\wh{\widehat}
\def\tr{\rm{Tr}}
\def\da{\dagger}
\newcommand{\beq}{\begin{equation}}
\newcommand{\eeq}{\end{equation}}
\newcommand{\beqa}{\begin{eqnarray}}
\newcommand{\eeqa}{\end{eqnarray}}
\newcommand{\intf}{\int_{-\infty}^\infty}
\newcommand{\into}{\int_0^\infty}

\begin{abstract}
The non-equilibrium dynamics of an ion chain in a highly anisotropic trap is studied when the transverse trap frequency is quenched across the value at which the chain undergoes a continuous phase transition from a linear to a zigzag structure. Within Landau theory, an equation for the order parameter, corresponding to the transverse size of the zigzag structure, is determined when the vibrational motion is damped via laser cooling. The number of structural defects produced during a linear quench of the transverse trapping frequency is predicted and verified numerically.
It is shown to obey the scaling predicted by the Kibble-Zurek mechanism, when extended to take into account the spatial inhomogeneities of the ion chain in a linear Paul trap.
\end{abstract}

\maketitle

The non-equilibrium statistical mechanics of long-range interacting systems is one of the challenging problem in statistical physics \cite{Gio}.
Nevertheless, close to a continuous phase transition it is sometimes possible to use concepts of equilibrium statistical mechanics in order to make some predictions for the system when the value of a control parameter is quenched through the critical value.
The Kibble-Zurek mechanism (KZM) has become a useful paradigm in this arena, accounting for a variety of phenomena ranging from the formation of massive particles in the early universe \cite{Kibble} to the vortex formation in superfluid Helium~\cite{Zurek}. The model applies to systems with a continuous phase transitions which is well described within Landau theory, and allows one to estimate the density of defects which are formed when quenching the control field $\nu_t$ across the critical value.
\indent In a nutshell, by comparing the characteristic time $\tau_Q$ of change of the control field $\nu_t$ with the relaxation time $\tau(\nu_t)$ of the system at equilibrium \cite{HHreview}, one identifies the corresponding freeze-out time scale, $\hat{t}$, which separates the regime in which the system follows adiabatically the quench, from the regime in which the system behaves as if the dynamics was frozen out. The correlation length $\xi$ at $\nu_t(\hat{t})$ then gives the characteristic length over which the system remains correlated, and hence the density of defects. The KZM prediction of the density of defects has been verified numerically \cite{kzmnum1,kzmnum2} and experimentally in a variety of systems~\cite{kzmexp}. Recently the model has been extended to describe the quench dynamics in a quantum phase transition~\cite{kzmqpt}.
The standard model of topological defects for homogeneous phase transitions breaks down whenever
the quench is local or the critical control parameter and the resulting transition become spatially dependent \cite{ikzm,Zurek09bec}.  This situation is arguable ubiquitous in nature and the main focus of this work.
In this Letter we study the out-of-equilibrium dynamics in a inhomogeneous laser-cooled Wigner crystal, quenched through its critical point.
% a structural continuous phase transition.
 %
 %%%%%%%%%%%%%%%%%%%%%%%%%%%%%%%%%%%%%%%%%%%%%%%%%%%%%%%%%%%%%%%%%%%%%%%%%%%%%%
\begin{figure}[t]
\includegraphics[width=8.0cm,angle=0]{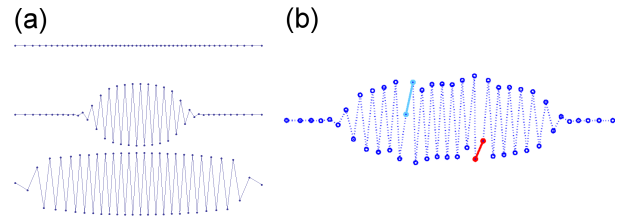}
\caption{\label{zztrap}
 (a) Charge distribution at equilibrium of an ion chain in a linear Paul trap for decreasing transverse trapping frequency (from top to bottom). Due to harmonic trapping, the density of ions is larger in the centre, where the repulsion is larger and the zigzag instability is first evident. (b) Charge distribution after a quench through the critical transverse frequency exhibiting both types of structural defects (solid lines): kink (light) and antikink (dark).
}
\end{figure}
%%%%%%%%%%%%%%%%%%%%%%%%%%%%%%%%%%%%%%%%%%%%%%%%%%%%%%%%%%%%%%%%%%%%%%%%%%%%%%
Here, we propose a novel system to test the KZM in a inhomogeneous phase transition which enjoys unprecedent level of experimental control and amenability for defect detection.
The Wigner crystal is composed of single-charged ions, which are confined in Paul or Penning traps and mutually repel via unscreened Coulomb interaction~\cite{DubinRMP}.
In highly anisotropic traps the ions can form a linear chain, which has a mechanical instability to a degenerate chain, with a zigzag structure, controlled by the density or by the transverse trap frequency \cite{fishman2008,RTSP08}. The form of the corresponding distribution of charges at equilibrium is shown in Fig.~\ref{zztrap}(a), for the case in which the ions are confined by a linear Paul trap. At equilibrium such instability is a second order phase transition~\cite{Schiffer93,Piacente2005,fishman2008}.
We determine the scaling of the density of defects after crossing such transition as a function of the cooling rate and of the quenching rate of the transverse trap frequency. In particular,
space like separated regions may develop zigzag structures with different orientations \cite{alex}, some of which are displayed in Fig.~\ref{zztrap}(b). These regions are the analogs of magnetic domains in a ferromagnetic material and the interface between domains is a structural  defect.

\indent The system we consider is composed by $N$ ions of mass $m$, charge $Q$ and coordinates ${\bm r}_n=(x_n,y_n,z_n)$ that are confined along the $x$-axis by a strongly anisotropic, radial trap. The Lagrangian describing the dynamics of the ions is $L=T-V$ where the kinetic and potential energies take the form, respectively, $T=\frac{1}{2} m\sum_n \dot{{\bm r}}_n^2$ and
$V=\frac{1}{2} m \sum_n [\nu^2x_n^2+\nu_t^2(y_n^2+  z_n^2)] +\sum_{n\neq n'} Q^2/(2|{\bm r}_n- {\bm r_n'}|)$, with $\nu$ and $\nu_t$ the frequency of the axial and the transverse confinement. At sufficiently low temperature and sufficiently large values of $\nut$ the ions form a chain along the $x$-axis.  In the local density approximation the linear density $n(x)$ is approximated by the function
$
n(x)=\frac{3}{4}\frac{N}{L}\left(1-\frac{x^2}{L^2}\right)
$
with $L$ the half-length of the chain and $x$ the distance from the center \cite{Dubin97}. The interparticle spacing  given by $a(x)=1/n(x)$ is a slowly-varying function of the position.
 In the thermodynamic limit, in which $a(0)\to a$ as the number of particles $N\to\infty$, one recovers in the center of the trap the statistical mechanics and dynamical properties of an infinite chain with uniform interparticle distance $a$~\cite{morigi2004,fishman2008}. The mechanical stability of the ion chain is warranted provided that $\nu_t>\nu_t^{(c)}$, such that the critical value $\nu_t^{(c)}$ is a function of the axial trap frequency and the ion density. At $\nu_t^{(c)}$ the chain undergoes a transition to a zigzag configuration, with transverse size $b$ \cite{Walther92}.

In the thermodynamic limit, the structural change is a second order phase transition, with $\nu_t$ a control field and $b$ the order parameter~\cite{fishman2008}. In particular, the critical value of the transverse frequency in the thermodynamic limit is given by $\nu_t^{(c)}=\omega_0\sqrt{7\zeta(3)/2}=(2.051\dots)\omega_0$, $\zeta$ being the Riemann-zeta function,
$\omega_0 =\sqrt{Q^2/ma^3}$ and $a$ the uniform interparticle spacing. For $N$ finite, the transition from a linear to a zigzag chain occurs at the value $\nu_t^{(c)}\approx 3N\nu/(4\sqrt{\log N})$, where the corrections scale with powers of $1/\log N$~\cite{morigi2004}, and corresponds to the relation $\nu_t^{(c)2}=4\frac{Q^2}{m a(0)^3}$ with $a(0)=1/n(0)$. For the system under harmonic axial confinement, the zigzag instability occurs first at the center of the trap, where the density is larger, and at lower values of $\nut$ extends towards the edge of the chain, as sketched in Fig.~\ref{zztrap}(a).

We now determine a Ginzburg-Landau (GL) equation for the position-dependent transverse size $b(x)$ of the zigzag chain, which is here a continuum field mode $\psi(x)$. The derivation is obtained assuming that a coarse-grained length scale $\delta x$ can be defined, with  $\delta x\gg a(x)$ and $a(x)\gg |\delta x ({\rm d} a(x)/{\rm d}x)|$. In this limit, we make a slowly-varying ansatz for the short-wavelength eigenmodes of the linear chain, such that for a given eigenmode we can write $\sigma_n=\alpha_n{\rm e}^{{\rm i}kna}$, with $\alpha_n$ slowly-varying amplitude~\cite{morigi2004}. Within the local-density approximation, we identify a local value of the critical transverse frequency
%\begin{equation}
$\nu_t^{(c)}(x)^2=4\frac{Q^2}{m a(x)^3}$
%\end{equation}
and write the Lagrangian $L=\int {\rm d}x{\mathcal L}(x)$ with the Lagrangian density
\begin{eqnarray}
\label{Eq:GL}
{\mathcal L}(x)&=&\frac{1}{2}\rho(x)\sum_{\sigma=y,z}\Bigl[(\partial_t\psi^{\sigma}(x))^2-h(x)^2\left(\partial_x\psi^{\sigma}(x)\right)^2 \nonumber\\
&&-\delta(x)\psi^{\sigma}(x)^2 -{\mathcal A}(x)\psi^{\sigma}(x)^4\Bigr]
\end{eqnarray}
where $\psi^{\sigma}(x)$ is the field, giving the zigzag size as a function of the position, $\rho(x)=mn(x)$ is the linear mass density, and  $\delta(x)=\nu_t^2-\nu_t^{(c)2}(x)$. The parameter $h(x)=\omega_0a(x)\sqrt{\log 2}$ is a velocity, and determines the speed with which a transverse perturbation propagates through the chain. Finally, the parameter ${\mathcal A}(x)=(93\zeta (5)/32)\omega_0^2/a(x)^2$ is positive and determines the value of the order parameter where $\delta(x)<0$. The Lagrangian density we derived has the form of a GL equation. It is valid for the modes of the linear chain close to instability and extends the theory presented in~\cite{fishman2008}.

The minimal energy solution of Eq.~\eqref{Eq:GL} fulfills the relation:
%\begin{equation}
%\label{eq:psi_infinite}
$\psi^{\sigma}( \delta(x) +2{\mathcal A}(x) [(\psi^{y})^2+(\psi^{z})^2]) = 0$,
%\end{equation}
 which always admits the solution $\psi^{\sigma} = 0$ corresponding to all the ions on the $x$ axis,
 and it is stable only for $ \delta(x)>0$ for all $x$, i.e. in the linear chain phase. For $ \delta(x)<0$ there is a continuous manifold of solutions of the form, $ \varrho(x)=\sqrt{- \delta(x)/2{\mathcal A}(x)}$, with $\varrho(x)=\sqrt{(\psi^{y})^2+(\psi^{z})^2}$, corresponding to the zigzag chain~\cite{fishman2008}. Within the Landau treatment~\cite{Landau}, one finds that the correlation function of the linear chain, evaluated for a static perturbation at a point, decays exponentially with the length scale $\xi\sim a\om_0/\sqrt{\delta(x)}$. In the following we focus on the situation in which the trap frequency in the $z$-axis is much larger than the $y$-axis so that $\psi(x)$ is along $y$ and we can drop the $\sigma$ label of the field.

Within the GL description we now assume that the transverse trap frequency $\nu_t$ undergoes a change in time in the interval $[-\tau_Q,\tau_Q]$, sweeping through the mechanical instability from the linear to the zigzag chain, such that $\nu_t^2=\nu_t^{(c)}(0)^2-\delta_0\frac{t}{\tau_Q}$ and $\nu_t^{(c)}(0)^2\gg\delta_0>0$. In this parameter regime, we can use the time-dependent parameter $\delta(x,t)=\nu_t^2(t)-\nu_t^{(c)}(x)^2$ inside the GL equation. We also assume that the chain is in contact with a thermal reservoir at low temperature $T$, which is warranted by laser cooling the chain motion. More specifically, we assume that some ions of the chain are Doppler cooled. In the Lamb-Dicke regime, where the mechanical effects of atom-photon interactions can be treated in perturbation theory, the energy distribution of the crystal modes obeys a Fokker-Planck equation~\cite{morigi2001}. The equation of motion for the field can be then written as
\begin{equation}
\label{Eq:field}
\partial_t^2\psi-h(x)^2\partial_x^2\psi+\eta \partial_t\psi+\delta(x,t)\psi+2{\mathcal A}(x)\psi^3=\varepsilon(t)
\end{equation}
where the scalar $\epsilon(t)$ is the Langevin force, describing the diffusion due to laser cooling, such that its moments fulfill the relations $\langle\varepsilon(  t)\rangle=0$, $\langle \varepsilon(   t) \varepsilon( t')\rangle=2\eta\kappa_BT \delta(t-t')$, where $\kappa_B$ is Boltzmann constant and $T$ is the temperature of Doppler cooling~\cite{Footnote}. In deriving Eq.~(\ref{Eq:field}), we have neglected axial distortions of the charge density due to the value of $\nu_t$.

We now estimate defect formation following a quench in the tranverse trapping frequency \cite{note}.
The nucleation of defects in such scenario resembles the formation of solitons in a cigar-shaped Bose-Einstein
condensate recently discussed by Zurek \cite{Zurek09bec}.
As a result of the inhomogneous charge distribution in the system, the transverse frequency is quenched through the critical point  at different times along the chain, giving rise to a propagating front along the axis, whose coordinates $(x_F,t_F)$ satisfy $\delta(x_F,t_F)=0$. The front velocity, at which the instability propagates, can be found by taking the ratio between the characteristic length of the control parameter, $\left(\partial_x\delta(x,t)/\delta(x,t)\right)^{-1}$, and the characteristic time scale at which it changes, $\left(\partial_t\delta(x,t)/\delta(x,t)\right)^{-1}$, giving $v_F\sim  \frac{\partial_t \delta(x,t)}{\partial_x\delta(x,t)}$. For the spatial dependence of the local critical frequency  $\nu_c^2(x)=\nu_c^2(0)[1-(x/L)^2]^3$, the front velocity takes the form
 $v_F\sim\frac{\delta_0}{\tau_Q}\left|\frac{d\nu_c^2(x)}{dx}\right|_{x_F}^{-1}=\frac{L\delta_0}{6\nu_t^{(c)}(0)^2\tau_Q}\frac{1}{|X|(1-X^2)^2},$ with $X=x/L$. Whenever the transition is homogeneous, $v_F$ becomes infinite and the standard scenario of defect formation of Kibble-Zurek applies: the density of defects in this case is simply determined by the correlation length at the freeze-out time scale $\hat{t}$. Elsewhere, the sound velocity comes into play. In order to compute it, we first note that the relative frequency can be written with reference to $t_F$ as $\delta(x,t)=-\delta_0[t-t_F(x)]/\tau_Q$, where $t_F(x)=\tau_Q[\nu_t^{(c)}(0)^2-\nu_t^{(c)}(x)^2]/\delta_0$. One can find the time scale relative to $t_F$, $\hat{\mathrm{t}}$,  at which the dynamics stop being adiabatic by equating the time scale $\delta/\dot{\delta}$ to the relaxation time $\tau$.
Two regimes can be identified, which refer to the relation between the damping ratio and the value of $\delta$ at $\hat{t}$. The so-called overdamped regime \cite{kzmnum1} corresponds to the situation $\eta\gg\sqrt{\delta(0,\hat{t})}$. In this case, $\hat{\mathrm{t}}=(\eta\tau_Q/\delta_0)^{1/2}$, which sets the freezed-out correlation length $\hat{\xi}_x=a\om_0/\sqrt{|\delta(x,\hat{\mathrm{t}})|}=a\om_0(\eta\delta_0/\tau_Q)^{-1/4}$.
Then, the characteristic velocity of a perturbation becomes $\hat{v}_x=\hat{\xi}_x/\hat{\tau}_x=a\om_0(\delta_0/\eta^3\tau_Q)^{1/2}$.
The condition for kinks formation
reads
%\beqa
$\frac{v_F}{\hat{v}_x}=\mathcal{A}_{o}\frac{1}{|X|}(1-X^2)^{-2}>1$
%\eeqa
 with
%$X=x/L$ and
%\beqa
$\mathcal{A}_{o}=\frac{L}{6\nu_t^{(c)}(0)^2 a\om_0\xi_0}\left(\frac{\eta\delta_0}{\tau_Q}\right)^{\frac{3}{4}}$.
%\eeqa
One can estimate  the effective size of the chain $2\hat{X_*}$ where the homogeneous KZM applies, by setting $\frac{v_F}{\hat{v}_x}=1$, and assuming $X_*\ll 1$ whence it follows that $|\hat{X_*}|\simeq\mathcal{A}_o$. The density of kinks obeys then the relation
\beqa
\label{over}
d_{\rm o}\sim \frac{2|\hat{X_*}|}{\hat{\xi}}=\frac{L }{3\nu_t^{(c)}(0)^2a^2\om_0^2} \frac{\eta\delta_0 }{\tau_Q}.
\eeqa
Note that this leads to a stronger dependence on $\tau_Q$ than in the homogeneous case, where defects can nucleate all over the system, and $d_{\rm o}\sim \hat{\xi}^{-1}=\frac{1}{a}\frac{1}{\omega_0}\left(\frac{\delta_0\eta}{\tau_Q}\right)^{1/4}$.
By contrast, in the underdamped regime ($\tau_Q\ll\delta_0/\eta^3$), the relaxation time is independent of the dissipation and diverges as $\tau=1/\sqrt{|\delta(x,t)|}$, which leads to the freeze-out time scale $\hat{\mathrm{t}}=(\tau_Q/\delta_0)^{1/3}$.
At this time scale, the correlation length reads $\hat{\xi}_x=a\om_0(\tau_Q/\delta_0)^{1/3}$
leading to a uniform sound velocity $\hat{v}_x=\hat{\xi}_x/\hat{\tau}_z=a\om_0$.
The causality argument implies that
%\beqa
$\frac{v_F}{\hat{v}_x}=\mathcal{A}_{u}\frac{1}{|X|}(1-X^2)^{-2}>1$,
%\eeqa
in terms of the parameter
%\beqa
$\mathcal{A}_{u}=\frac{L}{6\nu_t^{(c)}(0)^2 a\om_0\xi_0}$.
%\eeqa
For the purpose of deriving a scaling of the density of defects, we assume that formation of kinks arises only in a small central region $X_*\ll1$ of approximate size $2|\hat{X_*}|\simeq\mathcal{A}_{u}$, so that
\beqa
\label{under}
d_{\rm u}\sim\frac{2|\hat{X_*}|}{\hat{\xi}}
=\frac{L}{3\nu_t^{(c)}(0)^2a^2\om_0^2}\left(\frac{\delta_0 }{\tau_Q}\right)^{4/3},
\eeqa
 which should be compared with the density of defects in the homogeneous case
$
d_{\rm u}\sim \hat{\xi}^{-1}=\frac{1}{a}\frac{1}{\omega_0}\left(\frac{\delta_0}{\tau_Q}\right)^{1/3}.
$
We shall refer to the mechanism above as the inhomogeneous KZM (IKZM), whose main prediction is the scaling in Eqs.
(\ref{over}) and (\ref{under}), in dramatic contrast with their homogeneous counterparts.
\indent To test the IKZM we  next consider the dynamics of the structural phase transition for ions
 in a linear Paul trap, whose transverse square-frequency is driven through a linear quench as above.
We consider that the phase transition is well described by a Langevin dynamics following \cite{kzmnum1,morigi2001}.
The ions  obey the set of coupled stochastic equations
$m\ddot{{\bf r}}_i+\partial_{{\bf r}_i}V(\{{\bf r}_i\},t)+m\eta \dot{{\bf r}}_i+\varepsilon (t)=0$, ($i=1,\dots,N$)
where ${\bf r}_i=(x_i,y_i)$, $\varepsilon(t)$ is Langevin force (whose amplitude we take $\varepsilon=0.05$ $ml_0\nu^2$ with
$l_0^3=Q^2/m\nu^2$), and $\eta$ is the damping constant.
At $t=0$ the chain is in the ground state,
%(for $\nu_y=\nu_i$),
with all the ions at the equilibrium position satisfying
 $\partial_{{\bf r}_i}V(\{{\bf r}_i\})=0$. The value of $\delta(\hat{\mathrm{t}})$ is chosen such that the equilibrium configuration is linear, but close to the critical frequency below which the ground state becomes doubly degenerated. The system is then driven through the transition. Typical defects are shown in Fig. \ref{zztrap}b, and come in two varieties,  $\mathbb{Z}_2$ kinks (of topological charge $\sigma=+1$) and anti-kinks ($\sigma=-1$). These defects resemble the nonmassive kinks of the Frenkel-Kontorova model with a transversal degree of freedom which can be described by an effective $\phi^4$ theory for the translational displacement \cite{FKM}. When defects appear near the edges of the chain, they might be lost in the linear part.
 %
 %%%%%%%%%%%%%%%%%%%%%%%%%%%%%%%%%%%%%%%%%%%%%%%%%%%%%%%%%%%%%%%%%%%%%%%%%%%%%%
\begin{figure}[t]
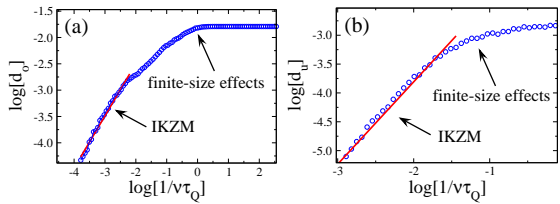

\includegraphics[width=3.6cm,angle=0]{ikzm_fig2.eps}
\includegraphics[width=3.6cm,angle=0]{ikzm_fig3.eps}
\caption{\label{IKZM}
Density of defects  for a harmonically trapped ion chain as a function of the inverse of the sweeping rate
(a) in the overdamped regime ($\eta=100\nu$), where the slope in the fit is $0.995$ with regression coefficient $1$.
(b) in the underdamped regime ($\eta=10\nu$), where the slope in the fit is $1.427$ with regression coefficient $0.994$.
The defects are only considered in the central $N_\mathcal{C}=30$ ions, in order to minimize defect losses
($N=50$, $2000$ realizations).
}
\end{figure}
%%%%%%%%%%%%%%%%%%%%%%%%%%%%%%%%%%%%%%%%%%%%%%%%%%%%%%%%%%%%%%%%%%%%%%%%%%%%%%
%
To minimise defects losses at the edges of the chain (see below), the density of defects $d$ (number of defects over the total number of ions) is computed once the average absolute transverse  displacement of the ions $\la y\ra=\sum_{i\in\mathcal{C}}|y_i|/N_\mathcal{C}$ approaches $90\%$ of that of the ground state in the final trap. $\mathcal{C}$ denotes the set of $N_\mathcal{C}$ central ions which would reach the zigzag structure in an adiabatic transition, and where the formation of defects is studied.
 The average density of defects $d$ over different realizations such as the one in Fig. \ref{zztrap}(b) is computed for different values of $\tau_Q$, and a least-squares fit to the list of data
 is used to extract the exponent governing the scaling. Numerical simulations in Fig. \ref{IKZM} are in good agreement with the IKZM scaling derived in Eqs. (\ref{over}) and (\ref{under}). Finite-size effects of the chain lead to a saturation of the density of defects and deviations from IKZM.
Further, the applicability of IKZM is restricted by the following effects:
a) Axial and transverse modes are coupled since the ions shift in the axial direction towards the center of the trap as the structural phase transition takes place.
b) The amplitude of the transverse displacement of the ions increases in the center of the trap, making the effective Peierls-Nabarro potential seen by a kink \cite{FKM} the convolution of a periodic potential with an inverted bell-shaped function, and leading to transport of defects and losses near the edges of the trap. Defect transport remains even if the longitudinal degrees of freedom of the ions are frozen on a lattice due to the transverse motion, and even when the trapping potential makes the inter-ion spacing homogeneous due to a local correction to the transverse critical frequency in the finite system.
%, $\nu_{t}^{(c)}(x_i)^2=\nu_t^{(2)}(0)^2-\sum_{n\neq n'}\frac{Q^2/m}{|x_n-x_{n'}|^3}$.
c) Defects with the same topological charge repel each other and attract otherwise. Scattering between kinks and anti-kinks can occur leading to their annihilation, a process particularly relevant in the underdamped regime which leads to deviations from the IKZM.

%{\it Conclusions}--
In conclusion, we have proposed an ion crystal as a test-bed for the formation of structural defects governed by the KZM. This system is far more amenable to experimental verification and control than other systems with realistic possibilities to enter the quantum regime. Though the paradigmatic result for an homogeneous second-order phase transition can be studied using a ring trap, the inhomogeneities in a linear Paul trap make an ion crystal a natural system to study the IKZM mechanism where the scaling of the number of topological defects as a function of the quenching rates is dramatically altered, as we have shown analytically and confirmed
by numerical simulations.

We thank T. Calarco, S. Fishman, and H. Rieger for fruitful discussions. Support by the European Commission (SCALA and QAP,
STREPs HIP and PICC), the EPSRC, and by the Spanish Ministerio de Educaci\'on y Ciencia (FIS2007-66944; FIS2008-01236;
Juan de la Cierva; Ramon-y-Cajal, Consolider Ingenio 2010 "QOIT").
G.M. and M.P. acknowledge the support of a Heisenberg Professorship and an Alexander-von-Humboldt Professorship, respectively.


\begin{thebibliography}{99}
\bibitem{Footnote} When the parameters of the laser drive are optimally chosen,
then $\kappa_BT=\hbar \gamma/2$, with $\gamma$ the radiative linewidth of the electronic transition~\cite{morigi2001}.
\bibitem{note} The quench can be realized by changing the ac component of the voltage of the electrodes.
For $N=50$, it is recquired to switch from $\nu_t=24 \nu$ to $12 \nu$ within $20 \frac{2\pi}{\nu}$ which should be realizable with current technology.
Higher switching rates  can be realized by adding a dc component in the radial direction and then changing it together with the dc in the axial direction in order not to alter the axial equilibrium position.
\end{thebibliography}
\end{document}